\documentclass[aps,prx,twocolumn,preprintnumbers,superscriptaddress]{revtex4} 

\usepackage{soul}
\usepackage[T1]{fontenc}
\usepackage{amstext,amsmath,amssymb,ulem,amsthm}
\usepackage{color}
\usepackage{float} 
\usepackage{times}
\usepackage{algorithm}
\floatname{algorithm}{Protocol}
\usepackage{ftnxtra}
\usepackage{enumitem}
\usepackage{graphicx}
\usepackage[colorlinks]{hyperref}
\usepackage[dvipsnames]{xcolor}

\usepackage[font=small,skip=0pt]{caption}

\allowdisplaybreaks[2]

\newcounter{thm}

\theoremstyle{plain}

\newtheorem{lemma}[thm]{Lemma}
\theoremstyle{definition}

\newtheorem{protocol}{Protocol}

\newcommand{\beq}{\begin{equation}}
\newcommand{\eeq}{\end{equation}}
\newcommand{\ket} [1] {\vert #1 \rangle}
\newcommand{\bra} [1] {\langle #1 \vert}

\newcommand{\Tr}{\mathop{\mathrm{Tr}}}

\newcommand{\ba}{\begin{align}}
\newcommand{\ea}{\end{align}}
\newcommand{\bea}{\begin{eqnarray}}
\newcommand{\eea}{\end{eqnarray}}

\makeatletter

\@ifundefined{textcolor}{}
{
 \definecolor{BLACK}{gray}{0}
 \definecolor{WHITE}{gray}{1}
 \definecolor{RED}{rgb}{1,0,0}
 \definecolor{GREEN}{rgb}{0,.6,0}
 \definecolor{BLUE}{rgb}{0,0,1}
 \definecolor{CYAN}{cmyk}{1,0,0,0}
 \definecolor{MAGENTA}{cmyk}{0,1,0,0}
 \definecolor{YELLOW}{cmyk}{0,0,1,0}
 }

\makeatother

\usepackage{bm}
\usepackage{mathtools}

\def\id{I}

\def\1{\mat{\id}}

\def\mat#1{\vec{#1}}

\renewcommand{\vec}[1]{\bm{\mathrm{#1}}}

\usepackage{cancel}

\begin{document}

\title{Client-friendly continuous-variable blind and verifiable quantum computing}

\date{\today}

\author{Nana Liu}
\email{nana.liu@quantumlah.org}
\affiliation{Centre for Quantum Technologies, National University of Singapore, 3 Science Drive 2, Singapore 117543}
\affiliation{Singapore University of Technology and Design, 8 Somapah Road, Singapore 487372}
\author{Tommaso F. Demarie}
\email{tommaso@entropicalabs.com}
\affiliation{Entropica Labs, 32 Carpenter Street, Singapore 059911}
\affiliation{Centre for Quantum Technologies, National University of Singapore, 3 Science Drive 2, Singapore 117543}
\affiliation{Singapore University of Technology and Design, 8 Somapah Road, Singapore 487372}
\author{Si-Hui Tan}
\affiliation{Centre for Quantum Technologies, National University of Singapore, 3 Science Drive 2, Singapore 117543}
\affiliation{Singapore University of Technology and Design, 8 Somapah Road, Singapore 487372}
\author{Leandro Aolita}
\affiliation{Instituto de F\'isica, Universidade Federal do Rio de Janeiro,
Caixa Postal 68528, 21941-972 Rio de Janeiro, RJ, Brazil}
 \affiliation{ICTP South American Institute for Fundamental Research, 
Instituto de F\'isica Te\'orica, UNESP-Universidade Estadual Paulista R. Dr. Bento T. Ferraz 271, Bl. II, S\~ao Paulo 01140-070, SP, Brazil}
\author{Joseph F. Fitzsimons}
\affiliation{Centre for Quantum Technologies, National University of Singapore, 3 Science Drive 2, Singapore 117543}
\affiliation{Singapore University of Technology and Design, 8 Somapah Road, Singapore 487372}

%%%%%%%%%%%%%%%%%%%%%%
\begin{abstract}
We present a verifiable and blind protocol for assisted universal quantum computing on continuous-variable (CV) platforms. This protocol is highly experimentally-friendly to the client, as it only requires Gaussian-operation capabilities from the latter. Moreover, the server is not required universal quantum-computational power either, its only function being to supply the client with copies of a single-mode non-Gaussian state. Universality is attained based on state-injection of the server's non-Gaussian supplies. The protocol is automatically blind because the non-Gaussian resource requested to the server is always the same, regardless of the specific computation. Verification, in turn, is possible thanks to an efficient non-Gaussian state fidelity test where we assume identical state preparation by the server. It is based on Gaussian measurements by the client on the injected states, which is potentially interesting on its own.
The division of quantum hardware between client and server assumed here is in agreement with the experimental constraints expected in realistic schemes for CV cloud quantum computing.
\end{abstract}
\maketitle 
Quantum computers promise  computational speedups for crucial classically-intractable problems. This includes the simulation of complex many-body quantum systems~\cite{lloyd1996universal, cirac2012goals, georgescu2014quantum}, searching through unstructured databases~\cite{Grover1997}, machine learning and artificial intelligence~\cite{harrow2009quantum, wiebe2012quantum, biamonte2017quantum,dunjko2017machine} and cryptography~\cite{shor1999polynomial, Lanyon2007}. However, scaling prototypical quantum processors to truly many-body regimes remains a technological challenge. Hence, similarly to the early classical computers, full quantum-computing capabilities are initially expected only at a few remote locations. Cloud quantum computing will then offer a means for clients to delegate their computations to a distant server with more powerful quantum hardware.

Delegating a computation, however, raises important security and privacy issues. This motivated the development of \emph{verifiable and blind assisted quantum computing}. Ideally, the client, Alice,  would like to delegate a computation to an untrusted server, Bob, while maintaining the privacy of her computation. At the same time, Alice needs a reliable certificate of the correctness of the computational output. The former property is known as \emph{blindness} and the latter as \emph{verifiability}~\cite{Fitzsimons2016}. Typically, the certificate is given by some form of test that the protocol must pass in order for its output to be accepted as valid by Alice. After the first proposals \cite{Childs2005b, aharonov2010proceedings, Broadbent2009}, which required repeated rounds of interaction between Alice and Bob, 
several improvements and variations followed~\cite{Aharonov2010,Fitzsimons2012,morimae2013blind, dunjko2012blind,Hajdusek2015, Gheorghiu2015, Reichardt2013, mckague2013interactive, kapourniotis2014verified, Broadbent2015,hayashi2015verifiable,morimae2016measurementonly,hayashi2016self, Mantri2016prx}. Importantly, preliminary experimental studies of assisted quantum computing have also been conducted~\cite{Barz2012, Barz2013, greganti2016demonstration, huang2017experimental}.

All of these developments have taken place in the qubit regime. In contrast, blind quantum computing on continuous-variable (CV) hardware is a much less explored territory \cite{Morimae2012b, marshall2016continuous}. To the best of our knowledge there is a single proposal reported~\cite{Morimae2012b} that allows the client to hide her input, output and her computation, whereas the scheme in \cite{marshall2016continuous} shows only the encryption of the input. CV degrees of freedom offer a competitive alternative to encode quantum information~\cite{Braunstein2005a, Andersen2010, Weedbrook:2012fe}, with some remarkable advantages over qubit-based platforms. For instance, CVs offer higher detection efficiencies and can be integrated into existing optical-fiber networks~\cite{Weedbrook:2012fe}, which are both highly desirable features for assisted computations. More generally, CV schemes have been explored in a variety of  settings~\cite{%Pati2000, Pati2002,
Marshall2015,liu2016power,Lau2017,Demarie:2014jx,Grosshans2002,menicucci2018anonymous,Douce2017}. Unfortunately, the seminal protocol of~\cite{Morimae2012b} displays an important intrinsic drawback for practical purposes. It puts a huge burden on Alice's shoulders in terms of experimental requirements and, in addition, requires repeated interaction between Alice and Bob. 

More precisely, the protocol of~\cite{Morimae2012b} requires that Alice performs single-mode {\it non-Gaussian} operations, while delegating the Gaussian entangling gates to Bob. Single-mode non-Gaussian operations are among the most experimentally challenging ones, commonly recognized as the main bottleneck for quantum computational universality in CV  platforms~\cite{Yukawa2013,Marshall2015-2, miyata2016implementation,marek2017general}. On the contrary, {\it Gaussian} operations --including maximally entangling gates-- are the most experimentally accessible ones for CV systems~\cite{Weedbrook:2012fe, Adesso2014}. They play a role analogous to Clifford operations in qubit systems. In fact, similarly to Clifford group operations on stabilizer states, any Gaussian CV computation can be efficiently simulated classically~\cite{Bartlett2002}. 
In contrast, any single non-Gaussian operation is enough to boost Gaussian quantum computations to universal ones~\cite{Lloyd1999}. Ironically, the situation for qubit systems is inverted: non-Clifford single-qubit gates are experimentally trivial, but Clifford entangling gates are not. Up to now, no CV scheme for blind quantum computing has been reported which is experimentally friendly to the client.

In this letter we fill this gap. We derive a verifiable and blind scheme for universal quantum computation on CV systems that requires only Gaussian quantum hardware on Alice's side. In addition, it requires neither repeated interaction between Alice and Bob nor universal quantum hardware on Bob's side. Bob's only requirement is the ability to prepare one kind of single-mode non-Gaussian state, e.g. the celebrated cubic phase state created by applying cubic phase gates~\cite{Yukawa2013,Marshall2015-2, miyata2016implementation,marek2017general} onto the vacuum. Our verification protocol is then based on the assumption that Bob is restricted to preparing identical copies of a resource state, which is the cubic phase state if Bob is honest. The difference in quantum hardware between Alice and Bob considered here reflects more fairly the actual constraints of real-life experiments. With this, our protocol lays the theoretical groundwork for realistic CV quantum cloud computing schemes.

\textit{Preliminaries}--In CV protocols, a single-mode quantum state is spanned by the Fock states denoted here by $\{\ket{n}, n=0,1,\ldots, \}$, where $n$ is an eigenvalue of the number operator $\hat{n}$. For a multimode CV state, let $\hat{x}_k$ and $\hat{p}_\ell$ be the position and momentum operators of the $k^{\text{th}}$ and $\ell^{\text{th}}$ modes respectively. These then satisfy the commutation relations $[\hat{x}_k, \hat{p}_\ell]=i\delta_{k,\ell}$. A quantum operation is said to be Gaussian when it is generated by a unitary $U=\exp(-iH)$, where the Hamiltonian $H$ is a second-order polynomial in the mode operators. An example is single-mode squeezing $S(s)=e^{i \log (s)(\hat{x}\hat{p}+\hat{p}\hat{x})}$ for $s\in\mathbb{R}$. Gaussian states are created by applying Gaussian operations onto the vacuum state. Gaussian measurements are an important subset of Gaussian operations and yield Gaussian distributed outcomes when applied to Gaussian states. These include homodyne detection which consists of the measurement of the quadrature $\hat{x}$ or $\hat{p}$ of a mode.

To implement an arbitrary CV computation, $U$, acting on an $m$-mode state $\ket{\Psi_{\text{in}}}$, one requires only the set of Gaussian operations, $\mathcal{G}$, including Gaussian measurements $\mathcal{M}$, and just one type of non-Gaussian operation~\cite{Lloyd1999}. Thus, $U$ can then be divided into sequences of Gaussian gates and non-Gaussian gates of the form $\openone_k\otimes C(\gamma) \otimes \openone_{m-k-1}$, where $0\leq k\leq m-1$. Here $\openone_k\equiv\openone^{\otimes k}$ where $\openone$ is the single-mode identity operator. 

We note that an example of a non-Gaussian operation that is needed for universality is the single-mode cubic phase gate 
$C(\gamma)=e^{i \gamma \hat{x}^3}$, where $\gamma \in \mathbb{R}$. When we apply the cubic phase gate to a finitely squeezed state $S(s)\ket{0}$, where $\ket{0}$ is the single-mode vacuum state, this gives rise to the following non-Gaussian state 
\begin{align}\label{eq:correctness1}
\ket{\tilde{\gamma}}_s=C(\tilde{\gamma})S(s)\ket{0}=\frac{e^{i \tilde{\gamma} \hat{x}^3}e^{-\hat{x}^2/(2s^2)}}{\sqrt{s}\pi^{1/4}}\int dx \ket{x}.  
\end{align}
We will later employ these as Bob's resource states for our assisted computation protocol. This is a finitely squeezed variant of the originally proposed cubic phase state~\cite{Gottesman2001}, the latter being less physical since it requires infinite squeezing as a resource. 

We now discuss  three important notions for an assisted computation protocol: \textit{correctness}, \textit{blindness} and \textit{verifiability}. 

\textbf{Definition 1.} Let $\ket{\Psi_{\text{out}}}$ denote the $m$-mode state that is the outcome of the intended computation that Alice wants to perform. Then $P_{\text{correct}}=\ket{\Psi_{\text{out}}}\bra{\Psi_{\text{out}}}$ is the projector onto the correct outcome of Alice's computation.  Let $\sigma_{\text{out}}$ be the outcome of Alice's computation when she delegates part of her computation to Bob \textit{and} Bob is honest. Then our delegation protocol is $\delta$-correct for $0\leq \delta \leq 1$ \footnote{Similar to definition in \cite{gheorghiu2017verification}.} when the probability of $\sigma_{\text{out}}$ being projected onto the correct outcomes satisfies
\begin{align} \label{eq:correctness}
\text{Tr}(P_{\text{correct}} \sigma_{\text{out}})\geq \delta \ .
\end{align}
This means that if Bob is honest, then with high probability Alice obtains the correct outcome to her computation if $\delta$ is large.

\textbf{Definition 2 (Blindness).} A delegation protocol is said to be \textit{blind} if the input state, the operations performed and the output state remain hidden from Bob (see \cite{Fitzsimons2016} and references therein for a formal definition). 

\textbf{Definition 3 ($\epsilon$-verifiability).} Suppose Alice requests quantum resource states from Bob to enable her to perform universal quantum computation. Let $\rho_{\text{out}}$ be the resulting outcome of this computation. The probability of $\rho_{\text{out}}$ projecting onto incorrect outcomes of the computation is denoted $\mathcal{P}(\text{incorrect})=\text{Tr}(P_{\text{incorrect}} \rho_{\text{out}})$, where $P_{\text{incorrect}}=\openone_m-\ket{\Psi_{\text{out}}}\bra{\Psi_{\text{out}}}$. Let $\mathcal{P}(\text{accept})$ be the probability that Alice \textit{accepts} the resource state given by Bob, according to her verification test. Then the assisted computation is said to be $\epsilon$-verifiable (for $0\leq \epsilon \leq 1$) if the joint probability $\mathcal{P}(\text{incorrect} \cap \text{accept}) \leq \epsilon$.

\textit{Blind delegation and verification protocol --} Alice wishes to perform an arbitrary CV quantum computation with output $U \ket{\Psi_{\text{in}}}$, where $U$ is a generic CV unitary operation and $\ket{\Psi_{\text{in}}}$ the $m$-mode Gaussian input state. Alice is only able to prepare Gaussian states, apply Gaussian gates and perform Gaussian measurements. To achieve universality in her computation, Alice delegates her non-Gaussian processing to Bob by requesting multiple copies of the non-Gaussian cubic phase state.  Alice only requests the same cubic phase state from Bob, so that he cannot infer any detail of the computation implemented by Alice. Thus, blindness is an intrinsic, built-in feature of the scheme and only an upper bound on the number of cubic phase gates in the computation is revealed to Bob. Verification, in turn, is based on a novel non-Gaussian state fidelity witness specially tailored for the cubic phase state, inspired by the witnesses of~\cite{aolita2015reliable}. This is measured by Alice on a subset of Bob's supplied states, used as test set. Remarkably, the witness requires only Gaussian measurements on at most four homodyne-detection bases per test mode, which is interesting in its own right. In addition, to estimate the expectation value of the witness, we use importance sampling techniques~\cite{gluza2017fidelity}, which allow the test-set size required for verifiability to scale only quadratically with the number of cubic phase states consumed by the computation. Hence, our protocol is not only experimentally friendly to Alice but also efficient in the number of single-mode non-Gaussian resource states required. We summarise our blind delegation and verification protocol below. 

\begin{protocol}\label{protocol:CVveri} Verified and blind assisted CV quantum computation~\\
\begin{enumerate}[start=0]
\item Alice's resources
 \begin{enumerate}
 \item A $m$-mode Gaussian state, $\ket{\Psi_{\rm in}}$ which is the input state for her computation.
 \item A circuit description representing Gaussian measurements and a unitary operation, $U$, that is decomposed into Gaussian gates and $M$ cubic phase gates.
 \item Parameters chosen for the verification test: threshold fidelity $F_T<1$, significance level $\beta$ (i.e., maximum failure probability of the test), and an estimation error $\eta$ that satisfies $\eta\leq (1-F_T)/2$.
 \end{enumerate}
 \item Alice requests $(N+1)$ copies of the pure state $\sigma=(\ket{\tilde{\gamma}}_s\bra{\tilde{\gamma}}_s)^{\otimes M}$ from Bob. We will see later how $N$ scales with $M$, $\beta$ and $\eta$. 
 \item Bob sends to Alice $(N+1)$ copies of an $M$-mode state $\rho$. If he is honest, $\rho=\sigma$. If Bob is dishonest, he sends Alice the state $\rho^{\otimes (N+1)}$ where $\rho \neq \sigma$ and we assume he cannot send more general states. 
 \item Alice retains the state $\rho$ for her computation and runs the verification test on the remaining $N$ copies of $\rho$. For the verification test, Alice makes an estimate $F_{\text{low}}^{\text{(est)}}$ of the quantity $F_{\text{low}} \equiv \text{Tr}(\mathcal{W}\rho)$. The observable $\mathcal{W}$ is a fidelity witness for the state $\sigma$, given in Eq.~\eqref{eq:Wdef}. The quantity $F_{\text{low}}$ is a lower bound on the fidelity $F(\sigma,\rho)$ between $\rho$ and $\sigma$. It can be estimated up to precision $\eta$ with homodyne detection on $\rho^{\otimes N}$, following the details of the importance sampling method in Appendix~\ref{app:sampling}. We say Alice \textit{rejects} $\rho^{\otimes N}$ if $F_{\text{low}}^{\text{(est)}}<F_T+\eta$ and \textit{accepts} otherwise. 
 \item If Alice \textit{accepts}, she uses the remaining state $\rho$ for her computation. More precisely, she uses $\rho$ to perform $M$ cubic phase gates on her input state $\ket{\Psi_{\text{in}}}$ by means of a gate teleportation protocol \footnote{See \cite{Ghose2007, Gottesman2001} for similar circuits.}. See Figs. 1 and 2. 
 \end{enumerate}	
\end{protocol}
When Bob is honest, gate teleportation and Gaussian operations allow Alice to approximately implement $C(\gamma)$ on any desired mode of her input state $\ket{\Psi_{\text{in}}}$. This protocol is both $\delta$-correct and blind as shown by the following theorem.
\begin{figure}[ht!]
\centering
\includegraphics[scale=0.4]{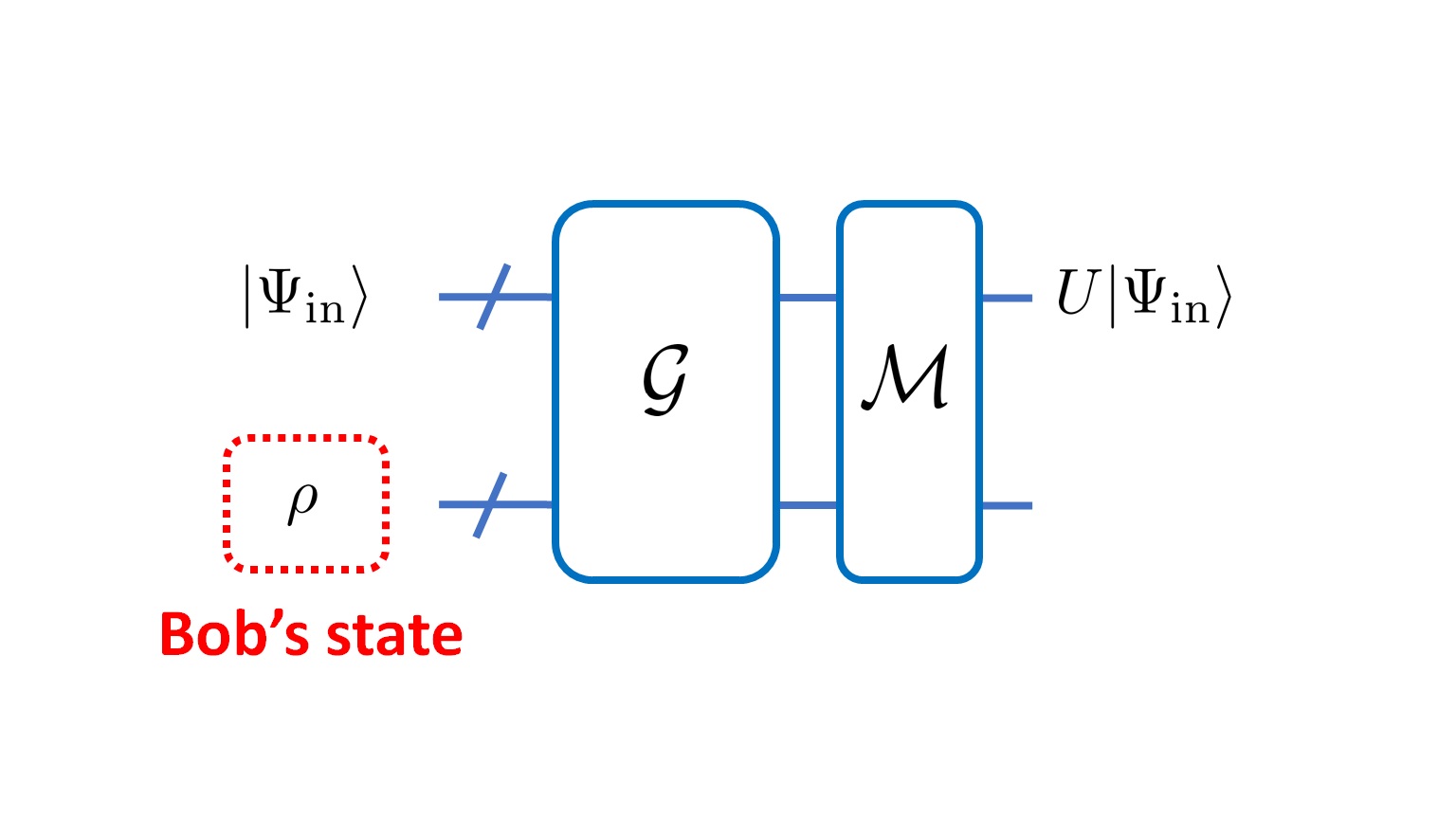}\label{fig:wholecircuit} 
\caption{\textit{Universal CV quantum computation.} To implement an arbitrary CV computation $U\ket{\Psi_{\text{in}}}$, one requires only Gaussian operations and at least one non-Gaussian operation. The non-Gaussian operation can be implemented by Alice when she uses Bob's non-Gaussian state resource $\rho$ and applies unitary Gaussian operations $\mathcal{G}$ and Gaussian measurements $\mathcal{M}$. See Fig.~2 and the text for more details.} 
\end{figure}

\textbf{Theorem 1.} Our assisted computation protocol is $\delta$-correct (where $\delta=1$) and reveals to Bob only an upper bound on the number of cubic phase states used.
  
\textit{Proof.} Our assisted computation protocol relies on the gate teleportation protocol in Fig.~2. If the $m$-mode input state in the top register is $\ket{\psi_{\text{in}}}$, then the $m$-mode output state in the top register is $\ket{\psi_{\text{out}}}_s=(\openone_{k} \otimes g_{s/r}(x_{\text{meas}}/r)C(\gamma) \otimes \openone_{m-k-1}) 
\ket{\psi_{\text{in}}}$, where $g_s(x_{\text{meas}})=e^{-(\hat{x}+x_{\text{meas}})^2/(2s^2)}$ and $x_{\text{meas}}$ is the measurement result in the bottom register. The gate teleportation protocol thus enables the application of a non-Gaussian operation on $\ket{\psi_{\text{in}}}$. More specifically, it applies a cubic phase gate on the $(k+1)^{\text{th}}$ mode of $\ket{\psi_{\text{in}}}$ up to a Gaussian factor for $k=0, 1,..., m-1$~\cite{Ghose2007, Gottesman2001}. For this protocol we can write $P_{\text{correct}}=\ket{\psi_{\text{out}}}_s \bra{\psi_{\text{out}}}_s$. Thus $\text{Tr}(P_{\text{correct}} \ket{\psi_{\text{out}}}_s\bra{\psi_{\text{out}}}_s)=1$ and we have perfect correctness, i.e., $\delta=1$ in Eq.~\ref{eq:correctness}. See Appendix~\ref{app:bqc} for more details. 

We note that if Bob's resource state is the infinitely squeezed version of the cubic phase state, the output state of Fig. 2 becomes exactly the cubic phase gate applied to the initial state $\ket{\psi_{\text{in}}}$, since $s\rightarrow \infty$ implies $g_s(x_{\text{meas}})\rightarrow 1$. Although finite $s$ gives a correction term to the cubic phase gate, this does not change our correctness argument since we only desire to perform a fixed non-Gaussian gate and not necessarily exactly the cubic phase gate. 

To show blindness in the sense that Bob can only learn the upper bound on the number of cubic states used, we first note that Bob has no access to any Gaussian part of the computation, which includes the input state $\ket{\psi_{\text{in}}}$ and the results of the (Gaussian) measurements. Furthermore, he cannot reconstruct the exact value of the parameters $\gamma$ used by Alice during the computation since Alice decides the squeezing parameter $r=(\gamma/\tilde{\gamma})^{1/3}$ used. This means the only useful information Bob obtains is the number of resource states that Alice requests, which is an upper bound on the size of the computation. \qed

\begin{figure}[ht!] 
\centering
\includegraphics[scale=0.4]{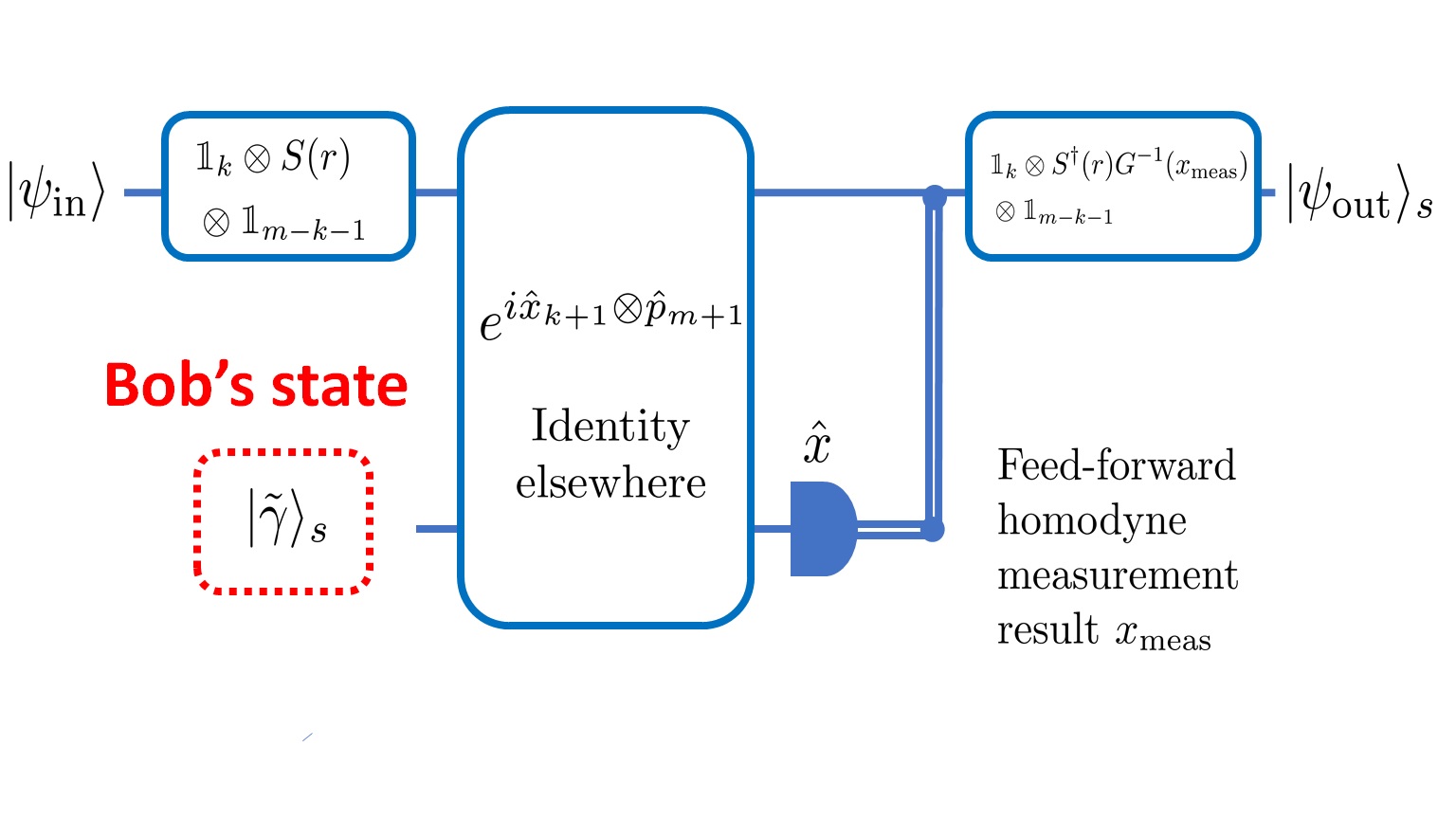} \label{fig:SUM}
\caption{\textit{Alice's circuit to implement a cubic phase gate.} Alice implements an approximation to the cubic phase gate $\openone_k\otimes C(\gamma)\otimes \openone_{m-k-1}$ acting on the state $\ket{\psi_{\text{in}}}$ by using the resource state $\ket{\tilde{\gamma}}_s$ given by Bob. Here $C(\gamma)$ acts on the $(k+1)^{\text{th}}$ mode of $\ket{\psi_{\text{in}}}$ and the resource state is on the $(m+1)^{\text{th}}$ mode. Here $S(r)$ is the single-mode squeezing operator with $r=(\gamma/\tilde{\gamma})^{1/3}$, where the value $r$ is only known to Alice. The initial gate $\openone_{k}\otimes S(r)\otimes \openone_{m-k-1}$ and final gate $\openone_{k} \otimes S^{\dagger}(r)G^{-1}(x_{\text{meas}})\otimes \openone_{m-k-1}$ acting on the top register are used to hide the value of $\gamma$ from Bob. Here $G^{-1}(x_{\text{meas}})=e^{-i \tilde{\gamma}x^3_{\text{meas}}}e^{-3i\tilde{\gamma}x_{\text{meas}}\hat{x}(\hat{x}+x_{\text{meas}})}$ is a Gaussian operator and $x_{\text{meas}}$ is the measurement outcome on the lower register of the operator $\hat{x}$.} 
\end{figure}
Now we describe Alice's verification test, which is based on the notion of fidelity witnesses \cite{Cramer2010,aolita2015reliable,Hangleiter2017,gluza2017fidelity}. These witnesses bypass the need for full state tomography \cite{lvovsky2009continuous} of Bob's state. They allow for verification in the general setting of independent and identically distributed (i.i.d) states and also non-i.i.d scenarios in the discrete-variable setting \cite{takeuchi2017verification}. In our scenario we assume that Alice has access to identical copies of Bob's resource state (the i.i.d setting). Our specific test relies on two ingredients. The first is to show that the fidelity between the final ideal and real $m$-mode states of Alice's computation is bounded from below by the fidelity between the ideal and real injected $M$-mode resource states. The second is to obtain a tight lower bound on the latter fidelity by means of a proper fidelity witness $\mathcal{W}$ using only Gaussian measurements, which Alice can perform. Finally, Alice's accept/reject decision is based on whether the value of the latter lower bound is high enough in comparison with the chosen threshold $F_T$. 

We start with the first ingredient. We call the initial ideal state that Alice possesses (on both input and injected modes) $\sigma_{\text{in}}=\ket{\Psi_{\text{in}}}\bra{\Psi_{\text{in}}} \otimes \sigma$, where $\sigma=(\ket{\tilde{\gamma}}_s\bra{\tilde{\gamma}}_s)^{\otimes M}$ is the ideal resource state. To perform a non-Gaussian operation on Alice's state $\ket{\Psi}_{\text{in}}$ in Fig.~1, the circuit in Fig.~2 is applied. Alice then applies Gaussian gates $\mathcal{G}$ and homodyne measurements $\mathcal{M}$ to her state. The outcome of Alice's intended computation consists of an $m$-mode pure state (for each measurement outcome) that can be expressed as $\sigma_{\text{out}}=\mathcal{E}(\sigma_{\text{in}})$, where $\mathcal{E}(\sigma_{\text{in}})=\text{Tr}_M(\mathcal{M}\mathcal{G} \sigma_{\text{in}} \mathcal{G}^{\dagger}\mathcal{M})/\text{Tr}(\mathcal{M}\mathcal{G} \sigma_{\text{in}} \mathcal{G}^{\dagger}\mathcal{M})$. Here $\text{Tr}_M$ denotes the partial trace over the $M$-mode system recording outcomes of her homodyne measurements. Bob actually provides Alice with $\rho$, so her real initial state is $\rho_{\rm in}=\ket{\Psi_{\text{in}}}\bra{\Psi_{\text{in}}} \otimes \rho$ and her output state would be $\rho_{\text{out}}=\mathcal{E}(\rho_{\rm in})$. Linearity of $\mathcal{E}$ implies the following lemma, proven in Appendix~\ref{app:rhoperp}.
\begin{lemma}
\label{lemma:fidelitybound} The fidelity between the final states $\sigma_{\text{out}}$ and $\rho_{\text{out}}$ satisfies the bound
\begin{align} \label{eq:FM}
F(\sigma_{\text{out}}, \rho_{\text{out}}) \geq F(\sigma_{\text{in}}, \rho_{\text{in}})=F(\sigma, \rho) \ .
\end{align}
\end{lemma}
We now focus on the second ingredient: efficient estimation the observable lower bound $F_{\text{low}}\equiv \text{Tr}(\mathcal{W}\rho)$ to the fidelity $F(\sigma,\rho)$ by measuring an adequate fidelity witness $\mathcal{W}$. A Hermitian observable $\mathcal{W}$ is a fidelity witness with respect to the target state $\sigma$ if it has the properties that 
\begin{align}
F_{\text{low}}\equiv \text{Tr}(\mathcal{W}\rho) \leq F(\sigma, \rho)
\end{align}
for all $\rho$ (universal lower bound) and $F_{\text{low}}=1$ for $\rho=\sigma$ (tightness). Our specific witness is given by the following: 
\begin{lemma}
The observable
\begin{align}\label{eq:Wdef}
&\mathcal{W}=\left(1+\frac{M}{2}\right)\openone_M-\sum_{k=0}^{M-1}\openone_k\otimes w_{k+1}\otimes \openone_{M-k-1} \ ,
\end{align}
is a fidelity witness with respect to the target state $\sigma$, where $w_{k+1}=(s^2/2)(\hat{x}_{k+1}^2+9\tilde{\gamma}^2 \hat{x}_{k+1}^4)+(1/(2s^2))(\hat{p}_{k+1}^2+2\tilde{\gamma}\hat{p}_{k+1}^3)+(1/(2s^2))\tilde{\gamma}((\hat{x}_{k+1}-\hat{p}_{k+1})^3-(\hat{x}_{k+1}+\hat{p}_{k+1})^3)$. Thus $\mathcal{W}$ is composed entirely of $\mathcal{O}(M)$ terms accessible through Gaussian measurements alone \footnote{For example, homodyne detection in quantum optics is sufficient to achieve this. Here the expectation value of a linear function of $\hat{x}$, $\hat{p}$ is related to the expectation value of the photon number difference detected in the two arms of a homodyne detection set-up. Thus higher powers of the expectation values of $\hat{x}$, $\hat{p}$ can be found by detecting the differences in the powers of the photon number operators. For example, see~\cite{Bachor2004}.} 
\end{lemma}
\textit{Proof.} See Appendix~\ref{app:fmeas}. \qed

Finally, we consider the accept/reject criterion of the verification test. A threshold fidelity $0<F_T<1$ and a significance level $0<\beta<1$ means that Alice must, with probability at least $1-\beta$, reject the state $\rho$ if $F(\sigma,\rho)<F_T$. To this end, the number $N$ of copies of $\rho$ she asks Bob for must be high enough for her to estimate $F_{\text{low}}$ up to precision $\eta$ and with failure probability at most $\beta$. In other words, the probability obeys $\mathcal{P}(|F_{\text{low}}^{\text{(est)}}-F_{\text{low}}|<\eta)\geq 1-\beta$. With this, she then rejects whenever $F_{\text{low}}^{\text{(est)}}<F_T+\eta$ and accepts otherwise. This guarantees the desired reject condition above. Conversely, if $\rho$ is accepted (i.e., if $F_{\text{low}}^{\text{(est)}}\geq F_T+\eta$), she knows that, with probability at least $1-\beta$, that $F(\sigma, \rho)\geq F_T$. 

The exact scaling of $N$ with respect to $M$, $\eta$ and $\beta$ defines the so-called \textit{sample complexity} of the test, which depends on the specific measurement scheme chosen. Here we use a scheme based on importance sampling \cite{gluza2017fidelity}. The basic idea of the method is to choose the observables to measure probabilistically according to their importance for $\mathcal{W}$. More precisely, Alice measures each single-mode quadrature operator appearing in the sum of Eq.~\eqref{eq:Wdef} with a probability proportional to the absolute value of the real factor in front of the operator. Hence, the relative importance of each observable dictates the frequency with which it is measured, with less important observables measured less frequently. This optimises the total number of measurements required. The details of the measurement procedure are described in Appendix~\ref{app:sampling}. With this, we obtain the following upper bound for $N$. 
\begin{lemma}
\label{lemma:samplingcom} (Sampling complexity of the verification protocol)~\\
If the number of copies of $\rho$ used in our verification test satisfies
\begin{align}
N\sim \mathcal{O}\left(\frac{M^2}{\eta^2} \ln \left(\frac{8}{\beta}\right)\right) \ ,
\end{align}
then $\mathcal{P}(|F^{\text{(est)}}_{\text{meas}}-F_{\text{low}}|<\eta) \geq 1-\beta$.
\end{lemma}	
\textit{Proof.}
We use a recent extension of Hoeffding's inequality for unbounded variables \cite{di2001probability} that leads to a sample complexity exponentially better in $\beta$ compared to previous scalings based on Chebyshev's inequality \cite{Aolita2015}. For details of the proof see Appendix~\ref{app:samplecomplexity}. 
\qed 

This result provides an upper bound to the sample complexity of our verification test and relies on the physical assumption of finite energy available per mode of $\rho$. This also makes $N$ efficient in the number of cubic phase states consumed by the computation.

Next, we shall show that our assisted protocol is $\epsilon$-verifiable, given the above results. 

\textbf{Theorem 2.} In Protocol \ref{protocol:CVveri}, when Bob is restricted to independent and identical preparation of the $M$-mode state $\rho$ and assuming finite energy available per mode of $\rho$, our assisted protocol is $\epsilon$-verifiable, where $\epsilon=1-(1-\beta)F_T$. 

\textit{Proof.} Let $\rho$ denote the resource state that Bob gives to Alice. Our aim is to bound $\mathcal{P}(\text{incorrect} \cap \text{accept})$, which is the probability that Alice accepts $\rho$ yet obtains an incorrect outcome to her computation.  From Bayes' rule and $\mathcal{P}(\text{accept})\leq 1$, we have $\mathcal{P}(\text{incorrect} \cap \text{accept})=\mathcal{P}(\text{incorrect}|\text{accept})\mathcal{P}(\text{accept})\leq \mathcal{P}(\text{incorrect}|\text{accept})$. Thus, to show $\epsilon$-verifiability, it suffices to find an upper bound for the conditional probability  $\mathcal{P}(\text{incorrect}| \text{accept})=\text{Tr}(P_{\text{incorrect}}\rho_{\text{out}})$, where $\rho_{\text{out}}=\mathcal{E}(\ket{\Psi_{\text{in}}}\bra{\Psi_{\text{in}}}\otimes \rho)$. Then applying Lemma \ref{lemma:fidelitybound} gives us
\begin{align}
\mathcal{P}(\text{incorrect}|\text{accept}) \leq 1-F(\sigma, \rho)\ . \label{eq:condineq}
\end{align}
Alice's accept condition implies that $F (\sigma,\rho) \geq F_T$ with probability at least $1-\beta$. %This is because $F_{\text{low}}^{\text{(est)}}(\sigma, \rho) \geq F_T+\eta$ and we know from lemma \ref{lemma:samplingcom} that the probability $|F_{\text{low}}^{\text{(est)}}(\sigma, \rho)-F_{\text{low}}(\sigma, \rho)| \leq \eta$ is at least $1-\beta$. 
This means we can now write $\rho=(1-\beta')(F' \sigma+(1-F')\sigma_{\perp})+\beta'\sigma'$, where $F'\geq F_T$, $\beta' \leq \beta$, $\text{Tr}(\sigma \sigma_{\perp})=0$ and $\sigma'$ is a quantum state. This implies $F(\sigma,\rho)=\text{Tr}(\sigma \rho) \geq (1-\beta')F' \geq (1-\beta)F_T$. Thus from Eq.~(\ref{eq:condineq}), we have
\begin{align}
\mathcal{P}(\text{incorrect}|\text{accept}) \leq 1-(1-\beta)F_T \ . \nonumber	
\end{align}
Choosing $\epsilon=1-(1-\beta)F_T$ gives us the bound we need.
\qed 

As a final remark, it is important to point out that the i.i.d assumption for Bob's state preparation can actually be removed. The quantum de Finetti theorem~\cite{konig2005finetti, renner2007symmetry} is a well-known and powerful tool to that end in discrete-variable scenarios. However, it is challenging for CV systems~\cite{d2007finite, renner2009finetti} because it requires the the number of registers (i.e. the sample complexity) to increase with the local Hilbert space dimension~\cite{MorimaeTH2017}. Fortunately, this can be overcome by adapting the recent techniques in~\cite{yuki}. This relies on Serfling's bound, which is an improvement over Hoeffding's bound as it does not require the i.i.d assumption (it considers sampling without replacement)~\cite{Serfling74}. Still, Serfling's bound requires the sampled variables to be bounded. Nevertheless, CV stabilizer states can be verified using a binary-outcome test based on the fact that they are extremal on stabilizer operators~\cite{yuki}. Since such a test defines a two-dimensional random variable, it can be handled with Serfling's bound. Remarkably, a similar test can be designed for the single-mode cubic phase state, as it is extremal on the fidelity witness $\mathcal{W}=3/2-w$ introduced in Eq.~\eqref{eq:Wdef}. More precisely, the cubic phase state is, by construction, a unique eigenstate of $\mathcal{W}$ with (maximal) eigenvalue 1. This allows us to safely relax the i.i.d assumption. We leave the details of this fascinating prospect for future work.

\section*{Acknowledgements}
The authors acknowledge support from the National Research Foundation and Ministry of Education, Singapore. LA's work is supported by the Brazilian agencies CNPq, CAPES, and FAPESP. The authors would like to thank Renato M. S. Farias for helpful comments on the manuscript. T.F.D. would like to thank Atul Mantri for interesting and valuable discussions about the concepts of blindness and verification. This material is based on research supported in part by the Singapore National Research Foundation under NRF Award No. NRF-NRFF2013-01, ANR-NRF grant NRF2017-NRF-ANR004 and the U.S. Air
Force Office of Scientific Research under AOARD grant FA2386-15-1-4082. 

\bibliographystyle{bibstyleNCM_papers}
\bibliography{allrefs-Tom}

\appendix 
\section{Gate teleportation protocol for the cubic phase gate} 
\label{app:bqc}
We begin with the circuit in Fig.~2 with initial state $(\openone_{m-1}\otimes S(r)\otimes \openone)\ket{\Psi}_{\text{in}} \otimes \ket{\tilde{\gamma}}_s$, where we choose $k=m-1$ here for simplicity. The results generalise easily for any other $k=0,..,m$.  Let ${\bf x}=(x_1,\ldots, x_m)$. We can write the $m$-mode state as $\ket{\Psi_{\text{in}}}=\int d^n{\bf x} \psi({\bf x})\ket{\bf x}$, for some bounded function $\psi({\bf x})$, then $(\openone_{m-1}\otimes S(r))\ket{\Psi_{\text{in}}}=\int d^m {\bf x} \psi_r({\bf x})\ket{\bf x}$. We apply the control operator $\openone_{m-1}\otimes \exp(i \hat{x} \otimes \hat{p})$ on the iniital state, and measure $\hat{x}$ in the last register with outcome $x_{\text{meas}}$. The final state becomes
\begin{align}
&\ket{\Psi}_s\otimes \ket{x_{\text{meas}}}\\
=&(\openone_m\otimes \ket{x_{\text{meas}}}\bra{x_{\text{meas}}})(\openone_{m-1}\otimes e^{i \hat{x} \otimes \hat{p}}) \nonumber\\
& \times \nonumber (\openone_{m-1}\otimes S(r) \otimes \openone)(\ket{\Psi_{\text{in}}} \otimes \ket{\tilde{\gamma}}_s) \nonumber \\
=&\frac{\openone_{m}\otimes \ket{x_{\text{meas}}}\bra{x_{\text{meas}}}}{\sqrt{s}\pi^{1/4}}(\openone_{m-1}\otimes e^{i\hat{x}\otimes \hat{p}}) \int d^m {\bf x}\int dx \psi_r({\bf x}) \nonumber \\
& \times e^{i\tilde{\gamma}x^3}e^{-x^2/(2s^2)}\ket{{\bf x},x}\nonumber \\
&=\frac{\openone_{m}\otimes \ket{x_{\text{meas}}}\bra{x_{\text{meas}}}}{\sqrt{s}\pi^{1/4}}(\openone_{m}\otimes e^{ix_m \hat{p}}) \int d^m{\bf x} \int x \psi_r({\bf x})  \nonumber \\
& \times e^{i\tilde{\gamma}x^3}e^{-x^2/(2s^2)} \ket{{\bf x},x} \nonumber \\
&=\frac{\openone_m\otimes \ket{x_{\text{meas}}}\bra{x_{\text{meas}}}}{\sqrt{s}\pi^{1/4}} \int  d^m{\bf x}\int dx e^{i\tilde{\gamma}x^3}e^{-x^2/(2s^2)}\psi_r({\bf x})  \nonumber \\
& \times\ket{{\bf x},x-x_m} \nonumber \\
&=\frac{(\openone_{m-1}\otimes G(x_{\text{meas}})e^{i\tilde{\gamma}\hat{x}^3}g_s(x_{\text{meas}})S(r)\otimes \openone)\ket{\Psi_{\text{in}}}\otimes \ket{x_{\text{meas}}}}{\sqrt{s}\pi^{1/4}} \ ,
\end{align}
where $G(x_{\text{meas}}) \equiv \exp(i\tilde{\gamma}x_{\text{meas}}^3) \exp(3i\tilde{\gamma} x_{\text{meas}}\hat{x} (x_{\text{meas}}+\hat{x}))$ is a unitary Gaussian correction in the operator $\hat{x}$, and $g_s(x_{\text{meas}})=\exp(-(\hat{x}+x_{\text{meas}})^2/(2s^2))$ is a smearing operation that applies a Gaussian envelope, with width $\sim 1/s^2$ centered on $x_{\text{meas}}$, onto the state it acts upon.

Using $S^{\dagger}(r) \hat{x} S(r)=r \hat{x}$, we can rewrite the above state as $\ket{\Psi}_s=G(x_{\text{meas}})e^{i \tilde{\gamma}\hat{x}^3}S(r)\ket{\tilde{\Psi}_{\text{in}}}$, where $\ket{\tilde{\Psi}_{\text{in}}}=g_{s/r}(x_{\text{meas}}/r)\ket{\Psi_{\text{in}}}$ is now a Gaussian-smeared state where the Gaussian envelope has width $\sim s/r$ centered on $x_{\text{meas}}/r$. Note that this Gaussian envelope is of the same type that appears in the usual CV cluster state computation \cite{Menicucci2006}. 

Then Alice applies a unitary Gaussian $\openone^{\otimes (m-1)}\otimes S^{\dagger}(r) G^{-1}(x_{\text{meas}})$ onto $\ket{\Psi}_s$ to obtain 
\begin{align}
\ket{\Psi_{\text{out}}}_s=e^{i\gamma \hat{x}^3}g_{s/r}(x_{\text{meas}}/r)\ket{\Psi_{\text{in}}} \ ,
\end{align}
where $r=(\gamma/\tilde{\gamma})^{1/3}$. 

Note that in the infinite squeezing $s\rightarrow \infty$ limit, we obtain the exact cubic phase gate operation  $\ket{\Psi_{\text{out}}}_{s\rightarrow \infty}=e^{i\gamma \hat{x}^3}\ket{\Psi_{\text{in}}}$. 
\section{Derivation of a lower bound for fidelity $F(\sigma_{\text{out}}, \rho_{\text{out}})$}\label{app:rhoperp}
First, we show that, for any mixed state $\rho_{\text{in}}$ and any pure state $\sigma_{\text{in}}$, there exists a density matrix $\sigma^{\perp}$ such that 
\begin{align} \label{eq:fidelityrhoperp}
\rho_{\text{in}}=F(\sigma_{\text{in}}, \rho_{\text{in}}) \sigma_{\text{in}}+(1-F(\sigma_{\text{in}}, \rho_{\text{in}}))\sigma^{\perp}
\end{align}
and $F(\sigma^{\perp}, \sigma_{\text{in}})=0$. In our delegation protocol, $\sigma_{\text{in}}$ is an $m+M$-mode state $\ket{\Psi_{\text{in}}}\bra{\Psi_{\text{in}}} \otimes \sigma$, where $\sigma$ is a pure $M$-mode product state. Given that $F(\sigma_{\rm in}, \rho_{\rm in})={\rm Tr}(\sigma_{\rm in}\rho_{\rm in})$, we can interprete this fidelity to be the projection of $\rho_{\text{in}}$ onto the subspace spanned by $\sigma_{\text{in}}$. This is because the trace of the product of two matrices is a valid Hilbert-Schmidt inner product. All the other components of $\rho_{\text{in}}$ must be in the orthogonal subspace to $\sigma_{\text{in}}$, $\sigma^{\perp}$. Thus Eq.~\eqref{eq:fidelityrhoperp} must hold while satisfying $F(\sigma^{\perp}, \sigma_{\text{in}})=0$. \\

Next, we demonstrate $\sigma^{\perp}$ is a valid density matrix. There are two requirements: $\text{Tr}(\sigma^{\perp})=1$, and $\sigma^{\perp}$ is positive semidefinite. The first condition follows directly by taking the trace on both sides of Eq.~\eqref{eq:fidelityrhoperp}. To show the latter, we rewrite $\sigma^{\perp}=\mathcal{O} \rho_{\text{in}} \mathcal{O}^{\dagger}$, where $\mathcal{O}=(\openone_{m+M}-\sigma_{\text{in}})/\sqrt{1-F(\sigma_{\text{in}}, \rho_{\text{in}})}$, which we note satisfies the requisite $\text{Tr}(\sigma_{\text{in}}\sigma^{\perp})=0$. Since $\rho_{\text{in}}$ is positive semidefinite, it can be written as $\rho_{\text{in}}=A^{\dagger} A$, for some matrix $A$. Thus $\sigma^{\perp}$ is also positive semidefinite because we can write $\sigma^{\perp}=(A \mathcal{O}^{\dagger})^{\dagger}(A \mathcal{O}^{\dagger})$. \\

Recall that $\sigma_{\text{in}}=\ket{\Psi_{\text{in}}}\bra{\Psi_{\text{in}}}\otimes \sigma$, where  $\sigma$ is a pure state, and the actual initial state to be tested is $\rho_{\text{in}}=\ket{\Psi_{\text{in}}}\bra{\Psi_{\text{in}}}\otimes \rho$, where $\rho$ is in general a mixed state. Then, using these in Eq.~(\ref{eq:fidelityrhoperp}) gives us
\begin{align} \label{eq:rhoRi}
\rho_{\text{in}}=F(\sigma, \rho) \sigma_{\text{in}}+(1-F(\sigma, \rho)) \sigma^{\perp} \ ,
\end{align}
where $F(\sigma_{\text{in}}, \sigma^{\perp})=0$. Applying the linear operator $\mathcal{E}$ that represents the teleportation circuit to Eq.~\eqref{eq:rhoRi},
\begin{align}
\mathcal{E}(\rho_{\text{in}})=F(\sigma, \rho) \mathcal{E}(\sigma_{\text{in}})+(1-F(\sigma, \rho)) \mathcal{E}(\sigma^{\perp}) \ .
\end{align}
Since $\sigma_{\text{out}}$ is a pure state, we can write the fidelity between $\sigma_{\text{out}}$ and $\rho_{\text{out}}$ as
\begin{align}
&F(\sigma_{\text{out}}, \rho_{\text{out}})=\text{Tr}(\sigma_{\text{out}}\rho_{\text{out}}) \ .
\end{align}
The fidelity between the final states $\sigma_{\text{out}}$ and $\rho_{\text{out}}$ then satisfies the bound
\begin{align}
&F(\sigma_{\text{out}}, \rho_{\text{out}}) = \text{Tr}(\sigma_{\text{out}}\rho_{\text{out}})=\text{Tr}(\mathcal{E}(\sigma_{\text{in}}) \mathcal{E}(\rho_{\text{in}})) \nonumber \\
&=F(\sigma, \rho)+(1-F(\sigma, \rho)) \text{Tr}(\mathcal{E}(\sigma_{\text{in}}) \mathcal{E}(\sigma^{\perp})) \nonumber \\
& \geq F(\sigma, \rho) \ ,
\end{align}
where in the last line we used the fact that $1-F(\sigma, \rho) \geq 0$ and $\text{Tr}(\mathcal{E}(\sigma_{\text{in}}) \mathcal{E}(\sigma^{\perp})) \geq 0$ since $\mathcal{E}(\sigma_{\text{in}})$ is pure (i.e., $\Tr(\mathcal{E}(\sigma_{\text{in}}^2)) = 1$) and $\mathcal{E}(\sigma^{\perp})$ is positive semidefinite. 

%%%%%%%%%%%%%%%%%%%%%%%%%%%%%%%%%%%%%
\section{Deriving $F_{\text{low}}$}\label{app:fmeas}
We can write our ideal $M$-mode resource state as $\sigma=(\ket{\tilde{\gamma}}_s \bra{\tilde{\gamma}}_s)^{\otimes M}=V^{\otimes M} \ket{0}_M\bra{0}_M(V^{\dagger})^{\otimes M}$, where $V=C(\tilde{\gamma})S(s)$ and $\ket{0}_M$ is the $M$-mode vacuum state. This means we can rewrite the squared quantum fidelity as
\begin{align} \label{eq:mfidelity}
F(\sigma, \rho)=\text{Tr}(\sigma \rho)=\text{Tr}(\ket{0}_M\bra{0}_M((V^{\dagger})^{\otimes M}\rho V^{\otimes M})) \ .
\end{align}
To find a lower bound to this quantity, we first note that 
\begin{align} \label{eq:zerom}
&\ket{0}_M\bra{0}_M \geq \openone_M-\sum_{k=0}^{M-1}\openone_k\otimes \hat{n}_{k+1}\otimes \openone_{M-k-1} \ ,
\end{align} 
where $\hat{n}_k$ is the number operator acting on the $k^{\text{th}}$ mode. We can see this inequality by acting the left and right-hand side with the Fock states $\ket{n_1,...,n_M}$, where $n_1,...,n_M$ are non-negative integers. These Fock states form a complete eigenbasis. When using the Fock state $\ket{0}_M$, the inequality above becomes an equality. Otherwise, the inequality implies $0 \geq 1-(n_1+...+n_M)$, which always holds. 

Since $(V^{\dagger})^{\otimes M}\rho V^{\otimes M}$ is positive semidefinite, then Eqs.~\eqref{eq:mfidelity} and ~\eqref{eq:zerom} gives the lower bound to the fidelity
\begin{align}\label{eq:fmeasdef}
F(\sigma, \rho)\geq \text{Tr}(\mathcal{W}\rho) \equiv F_{\text{low}} \ ,
\end{align}
where the fidelity witness $\mathcal{W}$ is 
\begin{align} \label{eq:wdef}
\mathcal{W}=\openone_M-\sum_{k=0}^{M-1}\openone_k\otimes V^{\dagger} \hat{n}_{k+1} V\otimes \openone_{M-k-1} \ .
\end{align}
The implication of this simple relation is that by writing $V \hat{n}V^{\dagger}$ in terms of $\hat{x}$ and $\hat{p}$, we can find a lower bound on fidelity by just measuring those quadratures of a given state $\rho$ to find how close it is to our true cubic phase state. Note that this is a tight bound. This means if $\sigma=\rho$, then $F=1=F_{\text{low}}$. \\

To compute $F_{\text{low}}$, we find $V \hat{n} V^{\dagger}$ in terms of $\hat{x}$ and $\hat{p}$ by first using 
\begin{align}
&S(s) \hat{n} S(s)^{\dagger}=a^{\dagger}a(2\text{cosh}^2(\log(s))-1) \nonumber \\
&+\text{cosh}(\log(s))\text{sinh}(\log(s))(a^{\dagger}a^{\dagger}+aa)+\text{sinh}^2(\log(s))\openone \ ,
\end{align}
where number operator $\hat{n}=a^{\dagger}a$ can be defined in terms of the creation and annihilation operators $a^{\dagger}=(1/\sqrt{2})(\hat{x}-i\hat{p})$ and $a=(1/\sqrt{2})(\hat{x}+i\hat{p})$ respectively. By also using $\exp(i\tilde{\gamma} \hat{x}^3)a^{\dagger}\exp(-i\tilde{\gamma}\hat{x}^3)=(\exp(i\tilde{\gamma}\hat{x}^3)a\exp(-i\tilde{\gamma}\hat{x}^3))^{\dagger}=(1/\sqrt{2})(\hat{x}+3i\tilde{\gamma}\hat{x}^2-i\hat{p})$, we find
\begin{align} \label{eq:unu}
&V \hat{n} V^{\dagger} 
=-\frac{1}{2}\openone+\frac{s^2}{2}(\hat{x}^2+9\tilde{\gamma}^2 \hat{x}^4)+\frac{1}{2s^2}(\hat{p}^2-6\tilde{\gamma} \hat{x}\hat{p}\hat{x}) \nonumber \\
&=-\frac{1}{2}\openone+\frac{s^2}{2}(\hat{x}^2+9\tilde{\gamma}^2 \hat{x}^4)+\frac{1}{2s^2}(\hat{p}^2+2\tilde{\gamma}\hat{p}^3) \nonumber \\
&+\frac{1}{2s^2}\tilde{\gamma}((\hat{x}-\hat{p})^3-(\hat{x}+\hat{p})^3) \ ,
\end{align}
where we used $2\hat{x}\hat{p}\hat{x}=\hat{p}\hat{x}^2+\hat{x}^2\hat{p}$ in the first line. Inserting Eq.~\eqref{eq:unu} into Eq.~\eqref{eq:wdef} we can write
\begin{align}\label{eq:wdef2}
&\mathcal{W}=\left(1+\frac{M}{2}\right)\openone_M-\sum_{k=0}^{M-1}\openone_k\otimes w_{k+1}\otimes \openone_{M-k-1} \ ,
\end{align}
where $w_{k+1}=(s^2/2)(\hat{x}_{k+1}^2+9\tilde{\gamma}^2 \hat{x}_{k+1}^4)+(1/(2s^2))(\hat{p}_{k+1}^2+2\tilde{\gamma}\hat{p}_{k+1}^3)+(1/(2s^2))\tilde{\gamma}((\hat{x}_{k+1}-\hat{p}_{k+1})^3-(\hat{x}_{k+1}+\hat{p}_{k+1})^3)$. Then we can write $F_{\text{low}}$ as the sum 
\begin{align}\label{eq:flowf}
F_{\text{low}}=1+\frac{M}{2}+\sum_{i=0}^{6M} \lambda_i \text{Tr}(\hat{f}_i \rho) \ ,       
\end{align}
where $\lambda_i$ are real coefficients and $\hat{f}_i$ are tensor-products of Gaussian operators with unit coefficients obtained by inserting Eq.~\eqref{eq:unu} into Eqs.~\eqref{eq:fmeasdef} and ~\eqref{eq:wdef}. Thus $\lambda_{1+6k}=-s^2/2$, $\lambda_{2+6k}=-9\tilde{\gamma}^2s^2/2$, $\lambda_{3+6k}=-1/(2s^2)$, $\lambda_{4+6k}=-\tilde{\gamma}/s^2$, $\lambda_{5+6k}=-\tilde{\gamma}/(2s^2)$, $\lambda_{6+6k}=\tilde{\gamma}/(2s^2)$ and $\hat{f}_{1+6k}=\openone_k \otimes \hat{x}_{k+1}\otimes \openone_{M-k-1}$, $\hat{f}_{2+6k}=\openone_n \otimes \hat{x}^4_{k+1}\otimes \openone_{M-k-1}$, $\hat{f}_{3+6k}=\openone_n \otimes \hat{p}^2_{k+1}\otimes \openone_{M-k-1}$, $\hat{f}_{4+6k}=\openone_k \otimes \hat{p}^3_{k+1}\otimes \openone_{M-k-1}$, $\hat{f}_{5+6k}=\openone_n \otimes (\hat{x}_{k+1}-\hat{p}_{k+1})^3\otimes \openone_{M-k-1}$, $\hat{f}_{5+6k}=\openone_k \otimes (\hat{x}_{k+1}+\hat{p}_{k+1})^3\otimes \openone_{M-k-1}$, where $k=0, 1, 2,...$ with a maximum value of $M-1$. 
\section{Sampling method} \label{app:sampling}
One method of directly estimating $F_{\text{low}}$ is to use importance sampling techniques \cite{gluza2017fidelity, Flammia:2011kl, daSilva:2011ej}. In this method, the relative importance of each observable, given by the size of the coefficients $\lambda_i$, is taken into account and less important observables do not require as many resources to estimate. 

From Eq.~\eqref{eq:flowf} we defined $F_{\text{low}}=1+M/2+\sum_{i=0}^{6M} \lambda_i \text{Tr}(\hat{f}_i \rho)$. Since $M$ is known, we only need to estimate the quantity $\sum_{i=0}^{6M} \lambda_i \text{Tr}(\hat{f}_i \rho)$. We then define a random variable $\mathbf{F}$ which takes the values $F_{i,f} \equiv \sum_{j=0}^{6M}|\lambda_j| \text{sign}(\lambda_i) f$, where $f$ are the eigenvalues of the quadrature operators $\hat{f}_i=\int f \ket{f_i}\bra{f_i}df$. We can also define a probability density $p(i,f)=p(i)p(f|i)$ for $\mathbf{F}$, where $p(i)=|\lambda_i|/\sum_{j=0}^{6M} |\lambda_j|$. The conditional probability term $p(f|i)=\text{Tr}(\hat{P}_{i,f}\rho)$, where $\hat{P}_{i,f}$ is the projector onto the $f^{\text{th}}$ eigenvalue of the $i^{\text{th}}$ quadrature measurement, defined by $\hat{f}_i=\int df f \hat{P}_{i,f}$. This means we can rewrite $F_{\text{low}}=\sum_{i=0}^{6M} \int df p(i,f) F_{i,f}$, which can be shown in the following 
\begin{align}
&F_{\text{low}}=\sum_{i=0}^{6M} \lambda_i \text{Tr}(\hat{f}_i \rho) \nonumber \\
&=\sum_{i=0}^{6M} \frac{|\lambda_i|}{\sum_{j=0}^{6M} |\lambda_j|} \text{Tr}(\text{sign}(\lambda_i)\sum_{k=0}^{6M} |\lambda_k| \hat{f}_i \rho) \nonumber \\
&=\sum_{i=0}^{6M} \frac{|\lambda_i|}{\sum_{j=0}^{6M} |\lambda_j|}\text{Tr}(\text{sign}(\lambda_i)\sum_{k=0}^{6M} |\lambda_k| \int df f \hat{P}_{i,f} \rho) \nonumber \\
&=\sum_{i=0}^{6M} \int df \frac{|\lambda_i|}{\sum_{j=0}^{6M} |\lambda_j|}\text{Tr}(\hat{P}_{i,f} \rho)\sum_{k=0}^{6M} |\lambda_k|\text{sign}(\lambda_i)f \nonumber \\
&=\sum_{i=0}^{6M} \int df p(i,f) F_{i,f} \equiv \langle \mathbf{F} \rangle \ .
\end{align}
In this way, we can consider $F_{\text{low}}$ as the expectation value of the random variable $\mathbf{F}$ which takes on the values $F_{i,f}$ with probability $p(i,f)$. 

To sample from $\mathbf{F}$, we begin by sampling the index $i$ with probability $|\lambda_i|/(\sum_{j=0}^{6M} |\lambda_j|)$. Then given this $i$, we measure the Gaussian observable $\hat{f}_i$, which outputs value $f$ with probability $\text{Tr}(\hat{P}_{i,f} \rho)$. Thus the corresponding output $F_{i,f}$ can be sampled with probability $p(i,f)$. 

For the $k^{\text{th}}$ sampling trial, where $k=1,...,N$, let the value of the corresponding $F_{i,f}$ be denoted $F^{(f)}$. For each $k^{\text{th}}$ trial, a single copy of $\rho$ is consumed. We can then obtain the estimate $F_{\text{low}}^{\text{(est)}}=(1/N) \sum_{k=1}^{N} F^{(k)}$ by using $N$ copies of $\rho$. In the limit $N \rightarrow \infty$, $F_{\text{low}}^{\text{(est)}}$ will output the exact value  $F_{\text{low}}$.
\section{Sample complexity} \label{app:samplecomplexity}
We use a recent extension of Hoeffding's inequality for unbounded variables \cite{di2001probability}, which works for independent and identically distributed variables. This new bound (see Theorem 1 in \cite{di2001probability}) shows that, for bounded $\langle {\bf F}^2 \rangle$, the probability $|F_{\text{low}}^{\text{(est)}}-F_{\text{low}}| \geq \eta$ is true is upper bounded by
\begin{align} \label{eq:newbound}
\mathcal{P}(|F_{\text{low}}^{\text{(est)}}-F_{\text{low}}| \geq \eta) \leq 8 e^{-N \eta^2/(33 \langle {\bf F}^2\rangle)} \ .
\end{align}
Thus the minimal number of copies of $\rho$ required to ensure $\mathcal{P}(|F^{\text{(est)}}_{\text{meas}}-F_{\text{low}}|<\eta) \geq 1-\beta$ is $N\sim \mathcal{O}((\langle \mathbf{F}^2\rangle/\eta^2) \ln (8/\beta)))$. In the following, we derive an upper bound to $\langle \mathbf{F}^2\rangle \leq K M^2$, where $K$ is a bounded constant independent of $M$. 

We know from Appendix~\ref{app:sampling} that $\mathbf{F}$ is a random variable which takes value $F_{i,f}$ with probability $p(i,f)$. Thus we can write the expectation value of $\mathbf{F}^2$ as 
\begin{align} \label{eq:finiteE}
&\langle \mathbf{F}^2\rangle=\sum_{i=0}^{6M} \int df p(i,f) F_{i,f}^2\nonumber \\
&=\sum_{i=0}^{6M} \int df \sum_{j=0}^{6M} |\lambda_j| |\lambda_i| \text{Tr}(\hat{P}_{i,f} \rho)f^2 \nonumber \\
&=\sum_{j=0}^{6M} |\lambda_j| \sum_{i=0}^{6M} |\lambda_i| \text{Tr}(\hat{f}_i^2 \rho) \nonumber \\
&\leq \text{max}(\text{Tr}(\hat{f}_i^2 \rho)) \sum_{j=0}^{6M} |\lambda_j| \sum_{i=0}^{6M} |\lambda_i|  \nonumber \\
& \leq \text{max}(\text{Tr}(\hat{f}_i^2 \rho)) (6M)^2 \text{max}(|\lambda_i|) \ ,
\end{align}
where $\text{max}(\text{Tr}(\hat{f}_i^2 \rho))$ denotes an upper bound for $\text{Tr}(\hat{f}_i^2 \rho)$ given any $i$ and $\text{max}(|\lambda_i|)$ denotes the maximum $|\lambda_i|$ value attained for $i$. 

We note that $\lambda_i$ depends only on the squeezing $s$ and $\tilde{\gamma}$ and under physical assumptions of finite energy available to Alice and Bob, $|\lambda_i|$ is bounded from above and is independent of $m$. To bound $\text{Tr}(\hat{f}_i^2 \rho)$, we observe that $\hat{f}_i$ are all local quadrature operators polynomial in $\hat{x}$ and $\hat{p}$ up to order $4$. Since the operators are local, the maximum value of $\text{Tr}(\hat{f}_i^2 \rho)$ is not expected to depend on $M$.   For cases where the quadrature operators are linear, then an upper bound on $\text{Tr}(\hat{f}_i^2 \rho)$ corresponds to an upper bound in energy per mode of $\rho$. Otherwise, we assume finite upper bounds of the higher moments of the quadrature operators. Inserting this into the inequality in Eq.~\eqref{eq:finiteE}, we find
\begin{align}
\langle \mathbf{F}^2\rangle \leq K M^2 \ ,
\end{align}
where $K$ is a bounded constant independent of $M$. 
\end{document}